# The Performance Evaluation of IEEE 802.16 Physical Layer in the Basis of Bit Error Rate Considering Reference Channel Models

Arifa Ferdousi[1], Farhana Enam[2] and Sadeque Reza Khan[3]

[1]Department of Computer Science and Engineering,
Varendra University, Rajshahi, Bangladesh
`arifaferdousi@yahoo.com`

[2] Dept. of Information and communication Engineering, Rajshahi University, Bangladesh
`farhana_ice2008@yahoo.com`

[3]Dept. of Electronics and Communication Engineering, NITK, Surathkal, India
`sadeque_008@yahoo.com`

*ABSTRACT*

*Fixed Broadband Wireless Access is a promising technology which can offer high speed data rate from transmitting end to customer end which can offer high speed text, voice, and video data. IEEE 802.16 WirelessMAN is a standard that specifies medium access control layer and a set of PHY layer to fixed and mobile BWA in broad range of frequencies and it supports equipment manufacturers due to its robust performance in multipath environment. Consequently WiMAX forum has adopted this version to develop the network world wide. In this paper the performance of IEEE 802.16 OFDM PHY Layer has been investigated by using the simulation model in Matlab. The Stanford University Interim (SUI) channel models are selected for the performance evaluation of this standard. The Ideal Channel estimation is considered in this work and the performance evaluation is observed in the basis of BER.*

## 1. INTRODUCTION

Broadband wireless access (BWA) is about bringing the broadband experience to a wireless context, which offers users certain unique benefits and convenience. Thus it has emerged as a promising solution for last mile access technology to provide high speed internet access in the residential as well as small and medium sized enterprise sectors. At present, Digital Subscriber Line (DSL) technology, which delivers broadband over twisted pair telephone wires, and cable modem technology, which delivers over coaxial cable TV plant, are the predominant mass-market broadband access technologies. But the practical difficulties in deployment have prevented them from reaching many potential broadband internet customers. Many areas throughout the world are not under the broadband access facilities. Even many urban and sub-urban areas cannot be served by DSL connectivity. As, it can reach about only three miles from the central switching office. Again many older cable networks do not have return channel which will prevent to offer internet access and thus many commercial areas are often not covered by cable network. But the BWA can overcome these types of situations and difficulties. Because of its wireless nature, it can be faster to deploy and easier to scale and more flexible, thereby giving it the potential to serve those customers who are not served or not satisfied by their wired broadband alternatives [1].

There are two fundamentally broadband wireless services (BWA). The first type attempts to provide a set of services similar to that of traditional fixed-line broadband, can be thought of as alternative to DSL or cable modem. The second type of BWA is called mobile broadband, which offers the additional functionality of portability, nomad city and mobility.

IEEE 802.16 standard and its associated industry syndicate, Worldwide Interoperability for Microwave Access WiMAX forum promise to offer high data rate over large areas, to a large number of users where broadband is unavailable. This is the first industry-wide standard that can be used for the fixed wireless access with substantially higher bandwidth than more cellular networks [2]. Wireless broad band systems have been in use for many years, but the development of this standard enables economy of scale that can bring down the cost of equipment, ensure interoperability, and reduce investment risk for operators.

The first version of IEEE 802.16 standard operates in the 10-66 GHz frequency band and requires line of sight (LOS) towers. Later the standard extended its operation through different PHY specification to 2-11 GHz frequency band enabling the no line of sight (NLOS) communications, which requires the techniques that efficiently mitigate the impairment of fading and multipath [3]. Taking the advantage of the OFDM technique the PHY is able to provide robust broad band service in hostile wireless channel.

The OFDM based physical layer of the IEEE 802.16 standard has been standardized in close cooperation with European Telecommunications Standard Institute (ETSI) High Performance Metropolitan Area Network (HiperMAN) [4]. Thus the HiperMAN standard and the OFDM based physical layer of IEEE 802.16 are nearly identical. Both OFDM-based physical layers shall comply with each other and a global OFDM system should emerge [3]. The WiMAX forum certified products for BWA comply with the both standards.

The main objective of this research work is to implement and simulate the IEEE 802.16 OFDM physical layer using the Matlab 7 to have better understanding of the standard and the system performance. This involves studying, through simulation, the various PHY modulations, coding schemes in the form of bit error rate (BER).

## 2. WIRELESS MAN-OFDM PHY LAYER

The WiMAX forum has adopted the version of 256-point OFDM based air interface specification for the reasons such as lower peak to average ratio, faster fast Fourier transform (FFT) calculation and less stringent requirements for frequency synchronization compared to 2048-point Wireless MAN-OFDMA. The size of the FFT point determines the number of subcarriers. Among these 256 subcarriers, 192 are used for user data, 56 are used as null for guard band and 8 are used as pilot subcarriers for various estimation purposes. The PHY allows to accept variable Cyclic Prefix (CP) length of 8, 16, 32 or 64 depending on the expected channel delay spread.

## 3. SIMULATION METHODOLOGY

As the research goal is to evaluate the performance of the IEEE 802.16 OFDM PHY layer, so this task involves modeling of the physical layer and propagation model. The simulator is developed by using the Matlab 7. Before discussing the physical layer setup OFDM symbol parameter is defined.Here there are two types of symbol parameter. One is primitive and one is derived. These two parameters characterize the OFDM symbol completely. The later one can be derived from the former one because of fixed relation between them. Here, in MATLAB implementation the primitive parameters are calculated as '*IEEE80216params*' which can be accessed globally.
The used OFDM parameters are given in the following table 1:

Table 1: OFDM Symbol Parameters

| | Parameters | Value |
|---|---|---|
| Primitive | Nominal Channel Bandwidth, BW | 1.75 MHz |
| | Number of subcarrier used, $N_{used}$ | 200 |
| | Sampling Factor, n | 8/7 |
| | Ratio of guard time to useful symbol time, G | 1/4, 1/8, 1/16, 1/32 |
| Derived | $N_{FFT}$(Smallest power of 2 greater than $N_{used}$) | 255 |
| | Sampling factor $F_s$ | floor(n.BW/8000)×8000 |
| | Subcarrier spacing, $\Delta f$ | $F_s / N_{FFT}$ |
| | Useful Symbol Time, $T_b$ | $1/\Delta f$ |
| | CP Time, $T_g$ | $G. T_b$ |
| | OFDM Symbol Time, $T_s$ | $T_b + T_g$ |
| | Sampling Time | $T_b / N_{FFT}$ |

## 4. PHYSICAL LAYER SETUP [4]

Figure 1 shows the base band part of the implemented transmitter and receiver. It corresponds to the PHY layer of the IEEE 802.16-2004 WirelessMAN-OFDM air interface. In this setup only the mandatory features are implemented. Channel coding part consists of three steps; Randomization, Forward Error Correction (FEC) and interleaving. Again FEC is done through two phases; outer phase is Reed-Solomon (RS) and Convolutional Code (CC). At the receiving end the complementary operations are implemented in the reverse order at the channel coding. The complete channel encoding setup is shown in figure 2 and the corresponding decoding set is shown in figure 3.

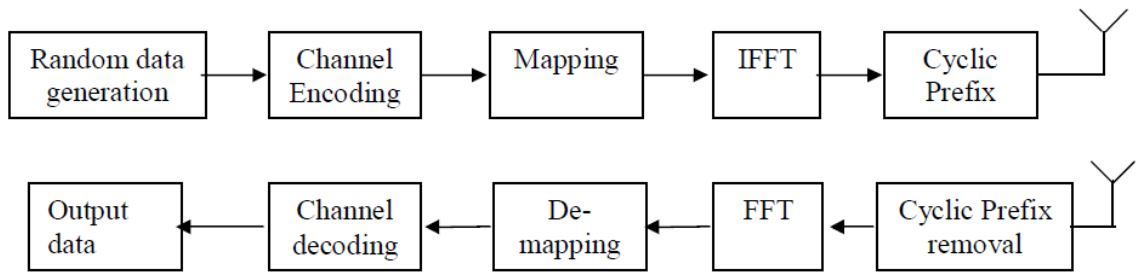

Fig: 1: Simulation Setup

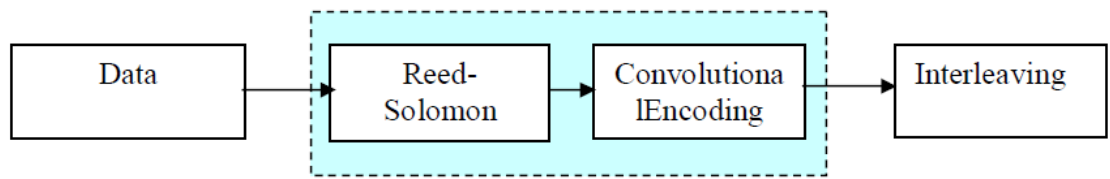

Fig: 2: Channel Encoding Setup

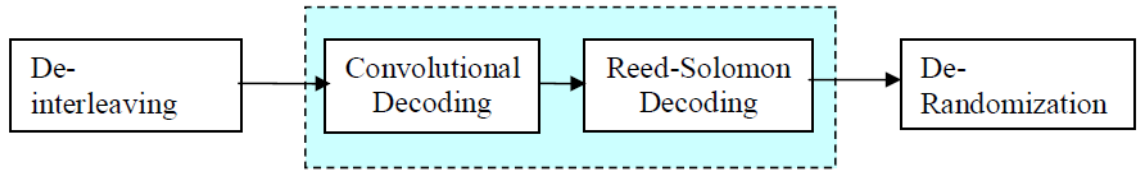

Fig: 3: Channel Decoding Setup

**Fig: 3: Channel Decoding Setup**

The Pseudo-Random Binary Sequence (PRBS) generator used for randomization is shown in figure 4. Each data byte to be transmitted enters sequentially into the randomizer, with the Most Significant Byte (MSB) first. Preambles are not randomized. The randomizer sequence is applied only to information bits.

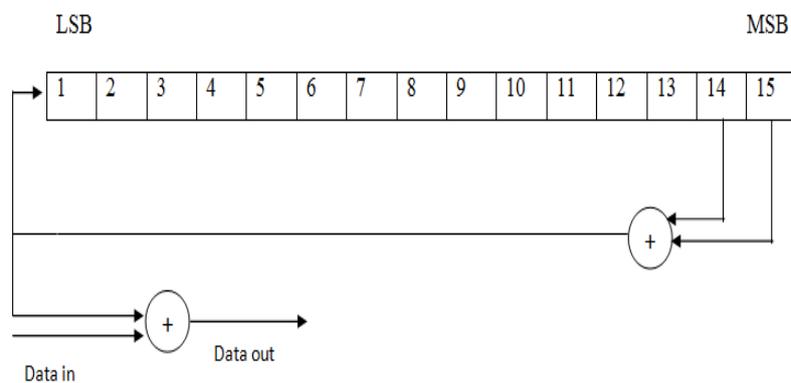

Fig: 4: PRBS generator used for data randomization in OFDM and OFDMA PHY. (From IEEE Std 802.16-2004.)

### 4.2. RS Encoder

The RS encoder of OFDM PHY is denoted as an $(N, K) = (255, 239)$ code, which is capable of correcting up to eight symbol errors per block. This Reed–Solomon encoding uses $GF(2^8)$, where GF is the Galois Field operator. The Reed–Solomon encoder and decoder require Galois field arithmetic. The following polynomials are used for the OFDM RS systematic code, an RS code that leaves the data unchanged before adding the parity bits:

$$G(x) = (x + \lambda^0)(x + \lambda^1)............(x + \lambda^{2T-1}), \lambda = 02_{HEX} \quad \text{............(1)}$$

$$p(x) = x^8 + x^4 + x^3 + x^2 + 1 \quad \text{............(2)}$$

The coding rate of the OFDM PHY RS encoder is then 239/255 (very close to one). The standard indicates that this code can be shortened and punctured to enable variable block sizes and variable error-correction capabilities.

### 4.3. CC Encoder

The outer RS encoded block is fed into inner binary convolutional encoder. The convolution code has an original coding rate of 1/2, has a constraint length of 7. In order to achieve variable code rate a puncturing operation is performed on the output of the convolutional encoder in accordance to table 2. In this table "1" denotes that the corresponding convolutional encoder output is used, while "0" denotes that the corresponding output is not used. At the receiver Viterbi decoder is used to decode the convolutional codes.

**Table 2**: Puncturing output configuration of the convolutional code.

| Rate | $d_{FREE}$ | X output | Y output | XY( punctured output) |
|------|------------|----------|----------|------------------------|
| 1/2  | 10         | 1        | 1        | X1Y1                   |
| 2/3  | 6          | 10       | 11       | X1Y1Y2                 |
| 3/4  | 5          | 101      | 110      | X1Y1Y2X3               |
| 5/6  | 4          | 10101    | 11010    | X1Y1Y2X3Y4X5           |

### 4.4. Interleaver

Interleaving is used to protect the transmission against long sequences of consecutive errors, which are very difficult to correct. The encoded data bits are interleaved by a block inter-leaver with a block size corresponding to the number of coded bits per allocated sub-channels per OFDM symbol. The Matlab implementation of the interleaver was performed calculating the index value of the bits after first and second permutation using the following equation 3 and 4 respectively

$$f_k = (N_{cbps}/12)k_{\mod 12} + floor(k/12) ....where, k = 0,1,2,...N_{cbps} - 1 \quad\quad (3)$$

$$s_k = s.floor(f_k / s) + (m + N_{cbps} - floor(12.m_k / N_{cbps}))_{\mod(s)}, where, k = 0,1,2,...N_{cbps} - 1 \quad\quad (4)$$

where s=ceil $(N_{cbps}/2)$, while $N_{cbps}$ stands for the number of coded bits per subscriber, i.e. ,1, 2, 3, 4 or 6 for BPSK, QPSK, 16-QAM OR 64-QAM, respectively.
The default number of subchannels i.e. 16 is used for this implementation.
The receiver also performs the reverse operation by following two permutation using equation 5 and 6

$$f_j = s.floor(j/s) + (j + floor(12.j / N_{cbps}))_{\mod(s)}, \quad j=0,1,2,\ldots N_{cbps}-1 \quad\quad (5)$$

$$s_j = 12.f_j - (N_{cbps} - 1).floor(12.f_j / N_{cbps}), \quad j=0,1,2,\ldots N_{cbps}-1 \quad\quad (6)$$

### 4.5. Constellation Mapper

The interleaved data are then entered serially to the constellation mapper. The Matlab implemented constellation mapper support BPSK, grey-mapped QPSK, 16-QAM and 64-QAM. The complex constellation points are normalized with the specified multiplying factor for different modulation scheme so that equal power is achieved for the symbols.

### 4.6. IFFT

The grey-mapped data are then sent to IFFT for time domain mapping. Mapping to time domain needs the application of the Inverse Fast Fourier Transform (IFFT). In this research the MATLAB

'ifft' function is called to do so. This block delivers a vector of 256 elements, where each complex number element represents one sample of the OFDM symbol.

### 4.7. Cyclic Prefix (CP) insertion

A cyclic prefix is added to time domain data to combat the effect of the multipath. Four different duration of cyclic prefix are available in the standard. The ratio of CP time to OFDM symbol time can be equal to 1/32, 1/6, 1/8, 1/4.

## 5. CHANNEL MODEL

To get a perfect evaluation of the performance of the developed communication system, an accurate description of the wireless channel is required to address its propagation, and the wireless channel is characterized by the factors like environment Path loss, Multipath delay spread, Fading characteristics, Doppler spread, Co-channel and adjacent channel interference [5]. While simulating the system these factors are considered. To take care of these requirement an empirical model is chosen, the Stanford University Interim (SUI) channel models. SUI channel models are an extension of the earlier work by AT&T Wireless and Erceg et al [6]. In this model a set of six channels was selected to address three different terrain types [7]. This model is used for simulation, design, and testing the technologies suitable for fixed broadband wireless applications [8]. The parameters for the model were selected based upon some statistical models. The parametric view of the six SUI channels is depicted in the following table 3 and 4.

Table 3: Terrain type for SUI channels

| Terrain Type | SUI channels |
|---|---|
| C (Mostly flat terrain with light tree densities) | SUI-1, SUI-2 |
| B (Hilly terrain with light tree density or flat terrain with moderate to heavy tree density) | SUI-3, SUI-4 |
| A (Hilly terrain with moderate to heavy tree density) | SUI-5, SUI-6 |

Table 4: General Characteristics of SUI Channels

| Doppler | Low delay spread | Moderate delay spread | High delay spread |
|---|---|---|---|
| Low | SUI-1, SUI-2 (High K-factor) |  | SUI-2 |
| High |  | SUI-4 | SUI-6 |

A scenario has been assumed with the following parameters:

- Cell size: 7 Km
- BTS antenna height: 30m
- Receive antenna height: 6m
- BTS antenna beam width: $120^0$
- Receive antenna beam width: omni directional
- Polarization: Vertical only.
- 90% cell coverage with 99.9% reliability at each location covered

International Journal on Cybernetics & Informatics ( IJCI) Vol.2, No.4, August 2013## 6. SUI CHANNEL MODEL IMPLEMENTATION

The goal of the model implementation is to simulate the channel coefficients. Channel coefficients with the specified distribution and spectral power density are generated using the method of filtered noise. A set of complex zero-mean Gaussian distributed number is generated with a variance of 0.5 for the real and imaginary part for each tap to achieve the total average power of this distribution is 1. In this way, a Rayleigh distribution (equivalent to Ricean with K=0) is got for the magnitude of the complex coefficients. In case of a Ricean distribution (K>0), a constant path component m has to be added to the Rayleigh set of coefficients. The K-factor specifies the ratio of powers between this constant part and the variable part. The distribution of the power is given below:

The total power P of each tap:
$$p = |m|^2 + \sigma^2 \quad \text{...............(7)}$$
Where $m$ is the complex constant and $\sigma^2$ is the variance of the complex Gaussian set
The ratio of the power is:
$$k = \frac{|m^2|}{\sigma^2} \quad \text{...............(8)}$$
From the above two equations the power of the complex Gaussian:
$$\sigma^2 = p \cdot \frac{1}{K+1} \quad \text{...............(9)}$$
and the power of the constant part as :
$$|m|^2 = p \cdot \frac{k}{K+1} \quad \text{...............(10)}$$
this SUI channel model address a specific power spectral density (PSD) function for the scatter component channel coefficients is given by:
$$S(f) = \begin{cases} 1 - 1.72 f_0^2 + 0.785 f_0^4, & |f_0| \leq 1 \\ 0, & |f_0| > 1 \end{cases} \quad \text{...............(11)}$$

where the function is parameterized by a maximum Doppler frequency $f_m$ and
$$f_0 = f/f_m \quad \text{...............(12)}$$
To generate a set of channel coefficients with this PSD function, original coefficients are correlated with a filter which amplitude frequency response is:
$$|H(f)| = \sqrt{S(f)} \quad \text{...............(13)}$$
For efficient implementation, a non recursive filter and frequency-domain overlap-add method has been used.

There are no frequency components higher than $f_m$ (for the construction formula of $S(f)$): So the channel can be represented with a minimum sampling frequency of $2 f_m$ according to the Nyquist theorem. For this reason the sampling frequency is chosen $2 f_m$. The power of the filter has to be normalized to 1, so that the total power of the output is equal to input one.

## 7. RESULT AND DISCUSSION

The simulator is developed in Matlab using modular approach. Each block of transmitter, receiver and channel is developed in separate 'm' file. The main procedure calls each block in a manner, so that a communication system works. The main procedure also contains the initialization parameters, input data, and it delivers results. The parameters like, number of OFDM symbols, CP length, modulation type, coding rate, range of SNR values and SUI channel models, can be set at the time of initialization. The input data stream is generated randomly. Output variables are available at the Matlab workspace. And the bit error rate (BER) for different SNR are stored in the text files to draw the plots. Each single block of transmitter is tested with its counter part of the receiver side to confirm that it works perfectly. Now the various BER vs. SNR plots are shown for all mandatory modulation and coding profiles as specified in the standard on same channel models. Figure 5, 6 and 7 show the performance on SUI-1, SUI-2, and SUI-3 channel models respectively. It is observed from the figures that the lower modulations and coding schemes provide better performance with less SNR. Larger distance between the adjacent points can tolerate larger noise at the cost of coding rate. Adaptive modulation schemes can be attain the highest transmission speed with a target BER. SNR required attaining at BER level $10^{-3}$ are listed in the following table 5.

Table 5: SNR Required at BER Level $10^{-3}$ for Different Modulation and Coding Profile

| Modulation | BPSK | QPSK | QPSK | 16-QAM | 16-QAM | 64-QAM | 64-QAM |
|---|---|---|---|---|---|---|---|
| CODE RATE | 1/2 | 1/2 | 3/4 | 1/2 | 3/4 | 2/3 | 3/4 |
| CHANNEL | SNR(dB) at BER level $10^{-3}$ | | | | | | |
| SUI-1 | 4.1 | 6.4 | 10 | 12.4 | 15.5 | 19.3 | 20.9 |
| SUI-2 | 7.4 | 10.4 | 14.1 | 16.20 | 19.5 | 23.2 | 25.5 |
| SUI-3 | 12.7 | 17.1 | 22.7 | 22.7 | 28.2 | 30 | 32.6 |

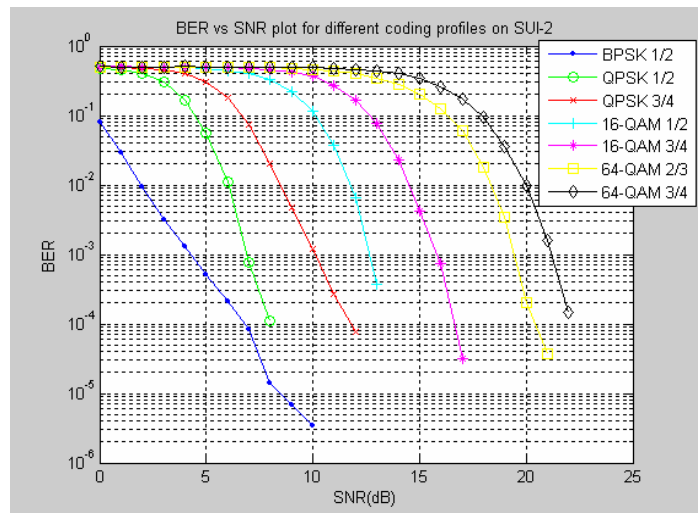

Figure 5: BER vs SNR plot for different coding profiles on SUI-1 channel.

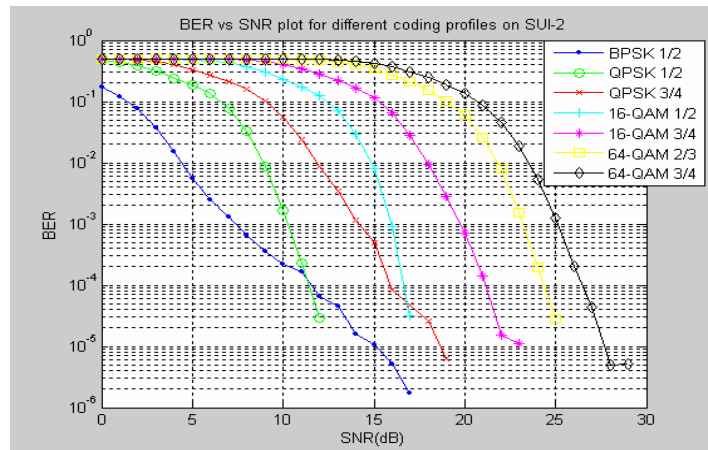

Figure 6: BER vs SNR plot for different coding profiles on SUI-2 channel.

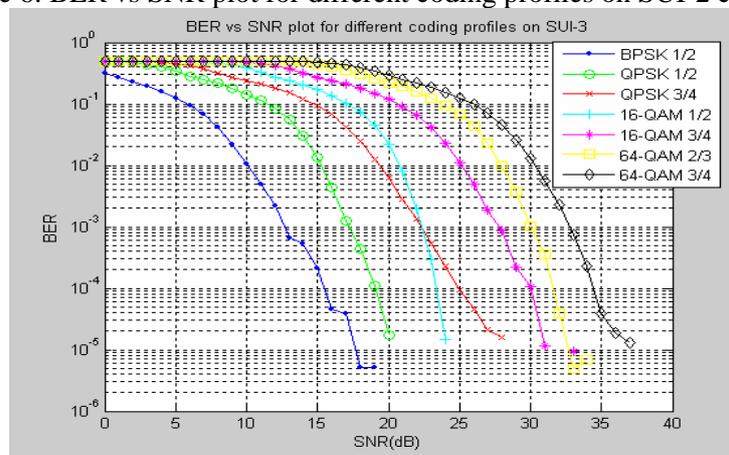

Figure 7: BER vs SNR plot for different coding profiles on SUI-3 channel.

## 8. CONCLUSION

The summery of this research is the implementation of the IEEE 802.16 OFDM PHY layer using Matlab in order to evaluate the PHY layer performance under reference channel model. The implemented PHY layer supports all the modulation schemes. To keep the research work doing over-sampling of the data is avoided before using the channel models. This can be implemented by minor modification. On the receiver side, perfect channel estimation was assumed to avoid the effect of any particular estimation method on the simulation results, though insertion of pilot subcarriers in the OFDM symbols make use of any comb-type estimator possible. In this performance evaluation, the effects of forward error correction (FEC) on different modulation under reference channel are not observed because of shortage of time. If it was done the performance evaluation will be completed perfectly.

Simulation is the methodology used to investigate the PHY layer performance. The performance evaluation method was mainly concentrated on the effect of channel coding on the PHY layer. The overall system performance was also evaluated under different channel conditions. Scatter plots were generated to validate the model in terms of general trends in reception quality as the different parameters were varied. A key performance measure of a wireless communication system is BER and BLER. Here the BER curves are used to compare the performance of different modulation and coding scheme used.

25

## Authors

Arifa Ferdousi received B.Sc. and M.Sc. degree in ICE from University of Rajshahi, Bangladesh, in the year of 2007 and 2009 respectively. Currently she is working as a lecturer in the department of CSE in Varendra University, Rajshahi, Bangladesh. Her research interest includes electronics system designing, OFDM, Advanced LTE Wi-Max and Bangla speech recognition system using Neural Network. She is the member of Bangladesh Electronic Society (BES).

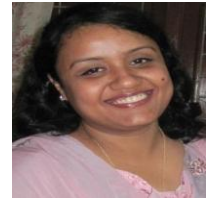

Sadeque Reza Khan received B.Sc. degree in Electronics and Telecommunication Engineering from University of Liberal Arts Bangladesh and continuing his M.Tech in VLSI from National Institute of Technology Kernataka (NITK), India. Currently he is in study leave from his Institution where he was working as a lecturer in the department of Electrical and Electronic Engineering in Prime University, Bangladesh. His research interest includes VLSI, Microelectronics, Control System Designing and Embedded System Designing.

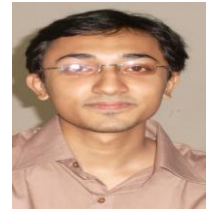